
\magnification=1200
\hsize=17 true cm
\vsize=22 true cm
\baselineskip=18 pt

%
%
%
%
\centerline{\bf Schwinger's Method for the Massive Casimir Effect}
\vskip 2.5 cm
\centerline{M. V. Cougo-Pinto, C. Farina and}
\medskip\noindent
\centerline{\it Universidade Federal do Rio de Janeiro}
\centerline{\it Instituto de F\'\i sica}
\centerline{\it Rio de Janeiro. RJ 21945-970, Brazil}
\centerline{A. J. Segu\'\i--Santonja}
\medskip\noindent
\centerline{\it Universidad de Zaragoza}
\centerline{\it Departamento de Fisica Teorica}
\centerline{\it Zaragoza - 50009, Spain}
\vskip 3 cm
\centerline{\bf Abstract}
\bigskip
We apply to the massive scalar field a method recently proposed by
Schwinger to calculate the Casimir Effect. The method is applied with
two different regularization schemes: the Schwinger's original one by
means of Poisson formula and another one by means of analytical
continuation.
\vfill
\eject
The importance of the Casimir Effect$^{(1)}$ steems from its far reaching
conceptual meaning in relativistic quantum field theory and for its
appearance in rather simple physical conditions, which makes possible not only
its precise calculation but also its experimental verification. The original
setup proposed by Casimir consists in a pair of paralel conducting plates
immersed in electrodynamic's vaccum and the predicted effect, the attraction
between the plates, was indeed later observed$^{(2)}$. The Casimir method of
calculation$^{(3)}$ (summation of the zero-point energies) as well as other
methods applied to several different setups, require the subtraction of
infinities to arrive at the final correct result. In such circumstances it
is highly desirable to have different methods of calculations, with the
infinities controled by different regularization schemes, in order to
check out the regularization independence of the result and further clarify the
conceptual meaning of the effect. For example, for the massless scalar field
in the original Casimir setup, an elegant  method recently proposed by
Schwinger$^{(3)}$ gives the force of atraction between the plates in the
context of source theory. Schwinger uses a regularization by means of
Poisson formula and subtracts infinities which are unrelated to the force
of atraction. We have shown in a previous letter$^{(4)}$ that a slight
modification in Schwinger's method, leads to the final result directly,
in an exceedingly short and simple way, without requiring any subtraction
of infinities. Our modification consists in the use of another
regularization prescription, by means of analytical continuation of
Schwinger's effective action. Here we further investigate Schwinger's
method  and its sensitivitness to the regularization procedure by applying
it to the more complicated case of a massive scalar
field in ($d+1$)-dimensional space-time. We use again the two different
regularization prescriptions: Schwinger's original one and
the other one by means of analytical continuation. The results
presented below show that the calculations by Schwinger's original method
are not much harder than in the massless case. On the other hand, in the
modified Schwinger's
method the calculations get a little bit more
involved. However, we should notice that some infinities to be subtracted
in the former case appear in the latter case as finite terms. Also, the
modified Schwinger's method is very simply related to other methods of
calculations, such as the zeta function method, as we will show elsewhere.
\par
Schwinger's method consists essentially in using Schwinger's formula for the
effective action $W^{(1)}$in proper time representation$^{(5)}$:
$$W^{(1)}(s_o)=-{i\over 2}\int_{s_o}^\infty {ds\over s}\,Tr\,e^{-isH}\,
+\,constant,\eqno(1)$$
where the exponential of $iW^{(1)}$ is defined as the vacuum persistency
probability amplitude $\langle 0_+\vert 0_-\rangle$ at the one-loop level,
$Tr$ means the total trace, $H$ is the corresponding proper time
Hamiltonian and $s_o$ is a regularization cutoff that must be sent to
zero only at the end of the calculations, after appropriate subtractions
of divergent terms is made; the additional constant is used to subtract
divergent terms, thereby establishing the physical normalization for
the situation. The energy of the configuration under consideration is
obtained from the prescription: ${\cal E}=-W^{(1)}/T$, where $T$ is the
duration of the measurement. Schwinger applied this method to calculate
the vaccum energy of the massless scalar field between two perfectly
conducting large parallel plates$^{(4)}$; the force between them
is then obtained as the derivative of the energy with respect to the plates
separation. A crucial part of the calculation is the use of Poisson's
formula$^{(6)}$,
%
%
$$\sum_{n=-\infty}^\infty\,e^{-n^2\pi\tau}=\tau^{-{1\over2}}\sum_{n=-\infty}^\infty
\,e^{-n^2\pi(1/\tau)},\eqno(2)$$
to exhibit two spurious terms in the energy, a uniform density vaccum
energy and the self-energy associated to each individual plate, both
diverging when $s_o\rightarrow 0$. To arrive at the final result these terms
must be subtracted before taking the limit $s_o\rightarrow 0$.
The same problem can be handled in a slightly modified way, by writing the
effective action as:
$$W^{(1)}(\nu)=-{i\over 2}\int_0^\infty {ds\over s}s^\nu\,Tr\,e^{-isH}\, ,
\eqno(3)$$
where, differently from (1), the range of integration
starts at zero and the factor $s^\nu$ in the integrand is taken large
enough to regularize the integral and disappears at
the end by analytical continuation to $\nu=0$. With this new regularization
scheme the final result for the massless case was obtained in a remarkable
simple way$^{(3)}$, without the appearance of any divergent term, actually
without the appearence of any spurious term.
\par
Let us now apply Schwinger's method to the proper-time Hamiltonian:
$$H=\sum_{j=1}^d\,p_j^2-\omega^2+m^2,\eqno(4)$$
where $p_j=-i\partial/\partial x^j$, $\omega=i\partial/\partial t$ and $m$ is
the mass. The boundary conditions are dictated by two perfectly conducting
hyperplanes, perpendicular to the $d$-direction and separated by a distance
$a$. The total trace is for this case:
$$Tr\,e^{-i\,s\,H}=T\,{i\,L^{d-1}\over(4\pi i)^{d/2}}{1\over s^{d/2}}\,
e^{-ism^2}\sum_{n=1}^\infty\,e^{-i\,s\,(\pi n/a)^2}.\eqno(5)$$
By substituing this trace in (1) we get for the energy ${\cal E}$ the
expression:
$${{\cal E}(s_o)\over L^{d-1}}=-{1\over 2}{1\over(4\pi i)^{d/2}}
\int_{s_o}^\infty{ds\over s^{1+(d/2)}}\,e^{-ism^2}\sum_{n=1}^\infty\,
e^{-is(\pi n/a)^2},\eqno(6)$$
which by using Poisson's formula can be recasted in the following form:
$$\eqalignno{{\cal E}(s_o)&=
L^{d-1}{1\over 4(4\pi i)^{d/2}}
\int_{s_o}^\infty{ds\over s^{1+d/2}}\,e^{-ism^2}\cr
&-aL^{d-1}{1\over 2(4\pi i)^{(d+1)/2}}
\int_{s_o}^\infty{ds\over s^{1+(d+1)/2}}\,e^{-ism^2}\cr
&-{1\over(4\pi i)^{(d+1)/2}}
\int_{s_o}^\infty{ds\over s^{1+(d+1)/2}}\,e^{-ism^2}\,aL^{d-1}
\sum_{n=1}^\infty\,e^{in^2a^2/s}.&(7)\cr}$$
The first term, on the right-hand side of this equation,
proportional to the conductor area $L^{d-1}$, comes from
the self-energy of each conductor and must be normalized to zero because
one is concerned only with the energy shift produced by varying the
distance $a$ between the conductors. The second term, proportional to the
spatial volume $aL^{d-1}$,
comes from a uniform spatial density of vaccum energy and must also
be eliminated by a term in the $constant$ of equation (1), in order
to normalize the vaccum energy density to zero in  the limit
$a\rightarrow \infty$.  Hence, we take the last term as the physical
energy of interaction; by taking the limit $s_o\rightarrow 0$
and changing the integration variable to $\sigma=a^2/is$ we arrive at:
$${{\cal E}(0)\over L^{d-1}}=-{1\over(4\pi)^{(d+1)/2}}
{1\over a^d}\sum_{n=1}^\infty \int_0^\infty d\sigma\,
\sigma^{{1\over 2}(d+1)-1}
\,e^{-n^2\sigma-m^2a^2/\sigma}.\eqno(8)$$
Since both $m^2a^2$ and $n^2$ are positive the above integral is well
defined and can be written as a series in modified Bessel
functions$^{(7)}$. The final expression for the energy
${\cal E}(0)$ is then obtained as$^{(8)}$:
$${{\cal E}(0)\over L^{d-1}}=-2\biggl({m\over 4\pi}\biggr)^
{(d+1)/2}{1\over a^{(d-1)/2}}\sum_{n=1}^\infty
{1\over n^{(d+1)/2}}K_{(d+1)/2}(2amn).\eqno(9)$$

Let us now turn to the modified Schwinger method, in which we start with
expression (3) and use the total trace (5) to arrive at the expression:
$${{\cal E}(\nu)\over L^{d-1}}=-{1\over 2}{1\over (4\pi i)^{d/2}}
\sum_{n=1}^\infty \int_0^\infty ds\,s^{(\nu-d/2)-1}\,
e^{-s[i\pi^2(n^2+a^2m^2/\pi^2)/a^2]}.\eqno(10)$$
The integration is now given by the mere definition of the gamma function,
$$\alpha^{-\xi}\Gamma(\xi)=\int_0^\infty ds\,s^{\xi-1}\,e^{-\alpha s},$$
and leads after some elementary manipulations to:
$${{\cal E}(\nu)\over L^{d-1}}=-{1\over 2} \biggl({\pi\over 4a^2}\biggr)^
{d/2} \biggl({a^2\over i\pi^2}\biggr)^\nu\,\Gamma(\nu-d/2)
\sum_{n=1}^\infty {1\over (n^2+\mu^2)^{\nu-d/2}},\eqno(11)$$
where we have defined $\mu=am/\pi$ and assumed $\nu>(d+1)/2$, to guarantee
the convergence of the series; its sum can then be analytically continued
by means of the following equality$^{(9,8)}$:
$$\sum_{n=1}^\infty {1\over (n^2+\mu^2)^z}=-{1\over 2\mu^{2z}}+{\sqrt{\pi}
\over 2\mu^{2z-1}\Gamma(z)}\biggl[\Gamma(z-1/2)+4\sum_{n=1}^\infty
(\pi n\mu)^{z-1/2}K_{z-1/2}(2\pi n\mu)\biggr],\eqno(12)$$
where $\Re z>1/2$. Using this equality in (11) we  arrive at:
$$\eqalignno{{{\cal E}(\nu)\over L^{d-1}}&={1\over 4}
\biggl({m\over 2\sqrt{\pi}}\biggr)^d{1\over(im^2)^\nu}
\Gamma(\nu-d/2)\cr
&-a{1\over 2}\biggl({m\over 2\sqrt{\pi}}\biggr)^{d+1}
{1\over(im^2)^\nu}\Gamma(\nu-(d+1)/2)\cr
&-2\biggl({m\over 4\pi}\biggr)^
{(d+1)/2}{1\over a^{(d-1)/2}} \biggl({a\over im}\biggr)^\nu
\sum_{n=1}^\infty
{1\over n^{(d+1)/2-\nu}}K_{(d+1)/2-\nu}(2amn).&(13)\cr}$$
Now we have to use the analytical continuation to take the limit
$\nu\rightarrow 0$ and this is all that we have to do in the massless case
to get the final result$^{(3)}$. However, in the present case of
a massive field we identify in (13) two spurious terms in the limit
$\nu\rightarrow  0$, not necessarily divergent as in (7),
but still spurious: the first term in (13),
independent of $a$, gives the self-energies associated with the individual
plates, and the second term, proportional to $a$, gives a uniform
energy density of the vaccum. After discarding these two terms and then
taking the limit we are left with an expression for the Casimir energy
which is identical to (9).
\par
After these calculations a short comment is in order. Schwinger's method,
which showed itself as a very powerful method for QED, is also a very
economical way of computing the Casimir energy in other situations.
Depending on the regularization scheme adopted, the spurious terms can
be made finite or divergent. In the regularization by Poisson formula
they are both divergent, while in the analytical continuation scheme
they will never be simultaneously divergent. It is interesting to note
that in the latter case the two terms exchange between themselves the
divergent character each time the dimension of space changes by a unit.
\par
The authors would like to thank M. Asorey and A. Tort for enlightning
discussions, and CNPq (The National Research Council of Brazil) and MEC-
CAICYT(Spain) for
partial financial support.One of us (A.J.S.S.) was also supported by
DGICYT (Spain), grant PB90-0916.
\bigskip
\noindent
{\bf References}
\medskip
\noindent
\item{1.} H. B. G. Casimir, Proc. K. Ned. Akad. Wet. {\bf 51} (1948),
793; for a review see G. Plunien, B. Muller and W. Greiner, Phys.
Rep. {\bf 134} (1986), 89.
\item{2.} M. J. Sparnay, Physica {\bf 24} (1958), 751.
\item{3.} J. Schwinger, Lett. Math. Phys. {\bf 24}, 59 (1992).
\item{4.} M. V. Cougo-Pinto, C. Farina, and A. J. Segu\'\i-Santoja,
Lett. Math. Phys., to be published.
\item{5.} J. Schwinger, Phys. Rev. {\bf 82} (1951), 664.
\item{6.} S. D. Poisson, Journal de l'Ecole Polytechnique, {\bf XII}
(cahier {\bf XIX}), (1823), 420.
\item{7.} Formula {\bf 3.471},9 in Gradshteyn, I. S. and Ryshik, I.
M., {\it Table of Integrals, Series and Products} (Academic Press,
New York and London, 1965).
\item{8.} J. Ambjorn and S. Wolfram, Annals of Phys. {\bf 147} (1983), 1.
\item{9.} E. Elizalde and A. Romeo, J. Math. Phys. {\bf 30} (1989),
1133; erratum, {\bf 31}, (1990), 771.
\bye